\documentclass[sigconf,natbib=false]{acmart}

\AtBeginDocument{%
  }

\setcopyright{acmcopyright}
\copyrightyear{2022}
\acmYear{2022}
\acmDOI{XXXXXXX.XXXXXXX}

\RequirePackage[
  datamodel=acmdatamodel,
  style=acmnumeric,
  ]{biblatex}

\usepackage{multirow}
\usepackage{algorithm}
\usepackage{algpseudocode}
\usepackage{bbold}
\usepackage{soul}
\usepackage{svg}
\DeclareUnicodeCharacter{0301}{\'{e}}
\begin{document}

\title{Personalized Promotion Decision Making Based on \\ Direct and Enduring Effect Predictions}


\author{Jie Yang}
\email{j-yang@mercari.com}
\affiliation{%
 \institution{Mercari, Inc.,}
 \city{Tokyo}
 \country{Japan}
 }

\author{Yilin Li}
\email{y-li@mercari.com}
\affiliation{%
 \institution{Mercari, Inc.,}
 \city{Tokyo}
 \country{Japan}
 }

\author{Deddy Jobson}
 \email{deddy@mercari.com}
\affiliation{%
 \institution{Mercari, Inc.,}
 \city{Tokyo}
 \country{Japan}
 }

\renewcommand{\shortauthors}{Yang et al.}

\begin{abstract}


Promotions have been trending in the e-commerce marketplace to build up customer relationships and guide customers towards the desired actions. Since incentives are effective to engage customers and customers have different preferences for different types of incentives, the demand for personalized promotion decision making is increasing over time. 

However, research on promotion decision making has focused specifically on purchase conversion during the promotion period (the direct effect), while generally disregarding the enduring effect in the post promotion period. To achieve a better lift return on investment (lift ROI) on the enduring effect of the promotion and improve customer retention and loyalty, we propose a framework of multiple treatment promotion decision making by modeling each customer's direct and enduring response. First, we propose a customer direct and enduring effect (CDEE) model which predicts the customer direct and enduring response. With the help of the predictions of the CDEE, we personalize incentive allocation to optimize the enduring effect while keeping the cost under the budget. To estimate the effect of decision making, we apply an unbiased evaluation approach of business metrics with randomized control trial (RCT) data. We compare our method with benchmarks using two promotions in Mercari and achieve significantly better results.


\end{abstract}




\maketitle

\section{Introduction} 

In the e-commerce marketplace, marketers run various types of promotions as a part of the customer relationship management (CRM) strategy in order to maintain customer relationships and guide consumers towards desired actions. Since incentives are an effective means to engage customers, and customers have different preferences for different types of incentives such as discount and rebate, multiple types of incentives are often used in promotions. In Mercari, Japan's largest C2C marketplace app, multiple treatment promotions are also a common marketing strategy. 





To better cater to each customer's needs, personalized promotion decision making is necessary and has garnered the attention of the research community. Some research optimizes promotion decision making by applying purchase prediction. Shen et al. \cite{shen2021framework} and Li et al. \cite{li2020spending} apply the purchase probability while Albert et al. \cite{albert2021commerce} use the Conditional Average Treatment Effect (CATE) on the purchase probability. While the modeling and optimization on the direct purchase probability brings profits to the business, there is still room for improvement. 

First, different customers contribute differently to the revenue even when they complete a purchase. According to the Pareto principle, roughly 80\% of consequences come from 20\% of causes. For promotions which aim to maximize revenue, we should consider not only the purchase probability but also the purchase amount in the decision making process.



Second, promotions can induce purchases but these often only turn out to be one-shot deals. On top of that, promotions might cause purchase acceleration or stockpiling. Therefore a post promotion dip in purchases is widely observed in marketing. This diminishes the effect of the promotion on customer loyalty \cite{ozer2012oxford}. To create promotions that aim to build long-term customer engagement and boost customer loyalty, the post-promotion period effect should be taken into account.

\begin{figure*}
  \centering
  \includegraphics[width=\textwidth]{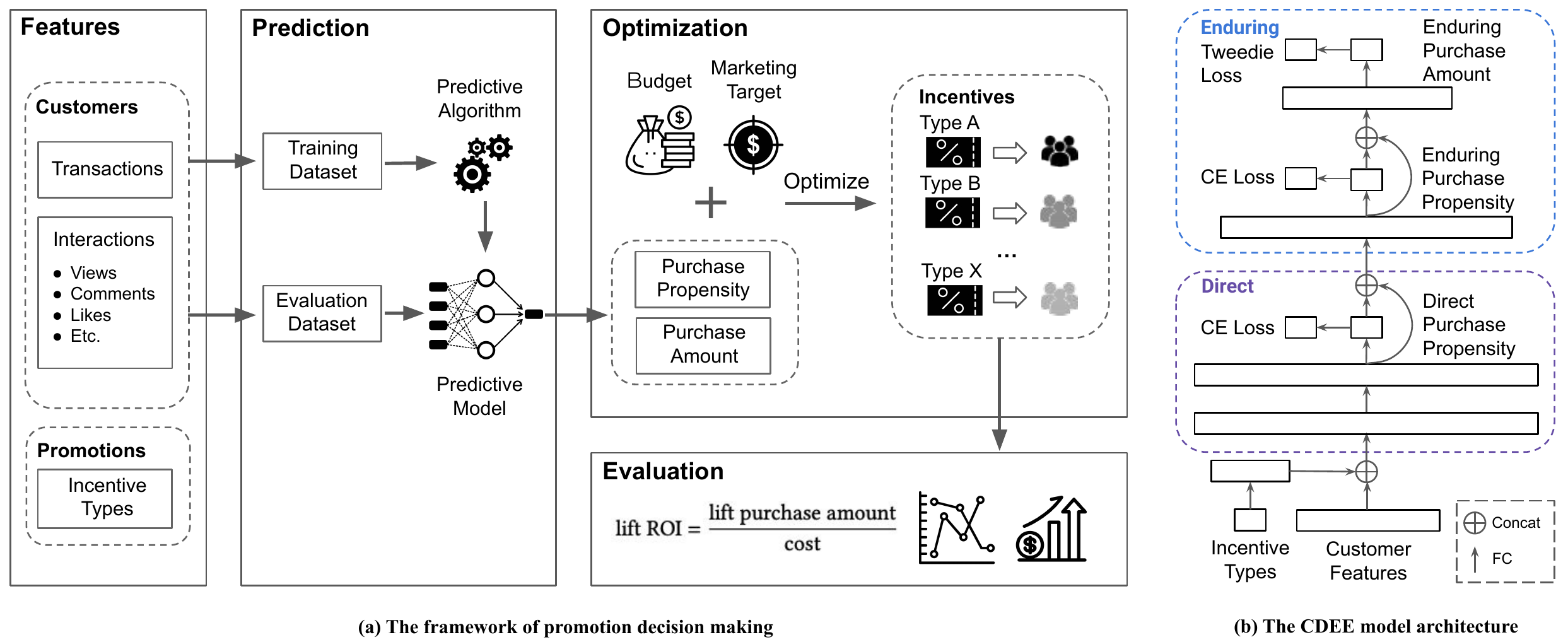}
  \caption{The framework of promotion decision making and the CDEE model architecture.}
     \label{fig:structure}
\end{figure*}



To target the efficient customers from the perspective of revenue during and post promotion, we propose a framework of promotion decision making by modeling customer direct and enduring response. The direct response corresponds to the customer's response during the promotion period, while the enduring response corresponds to the customer's response during and after the promotional period. Our framework consists of both direct and enduring response predictions, incentive allocation decision making and business impact evaluation. 




We propose a customer direct and enduring effect (CDEE) model which provides predictions of the direct purchase propensity and the enduring purchase amount of each customer. Since the distribution of the purchase amount is highly skewed, it makes purchase amount regression more difficult than purchase propensity prediction. To fit the distribution, we apply a regression loss derived from the compound Poisson–Gamma distribution \cite{tweedie1984index}, which models a positive mass at zero and a continuous response. Furthermore, we construct a multi-task learning structure to exploit commonalities across tasks of direct and enduring response learning, and purchase propensity and amount learning.

Referrig to as prescriptive analytics \cite{borges2021strategic}, we formalize the incentive allocation problem. Our optimization goal is to pick the best incentive types for customers maximizing the total enduring purchase amount while keeping the cost under budget.

To estimate the effect of decision making, we apply an unbiased evaluation approach on the business metrics with randomized control trial (RCT) data. To solve the overestimation issue of decision making which takes predictions as inputs \cite{ito2018unbiased}, we evaluate the experiments with cross-validation. We conduct comparative experiments on two promotional campaigns in Mercari.




We summarize our major contributions as follows:
\begin{itemize}
\item We propose a framework of promotion decision making by targeting customer direct and enduring effect. Our framework consists of direct and enduring response predictions, incentive allocation decision making and business impact evaluation. 
\item We propose a customer direct and enduring effect (CDEE) model, which provides predictions of the direct purchase propensity and the enduring purchase amount of each customer. 
\item We compare the CDEE model against benchmarks on two promotional campaigns in our company and achieve a significantly better performance.
\end{itemize}

This paper is organized as follows. Section 2 provides an overview of related work regarding prescriptive analytics and customer behavioral characteristic prediction. Section 3 illustrates the framework, and further describes the optimization problem formulation and the evaluation approach in detail. Section 4 elaborates on the architecture of the CDEE model. Section 5 represents the dataset and experiment results. Section 6 summarizes the paper.


\section{Related Work}

\subsection{Prescriptive Analytics}

Prescriptive analytics, which seeks to find the best course of action for business decisions, has been increasingly gathering research interest. It aims at suggesting (prescribing) the best decision options by incorporating the predictive analytics output and utilizing artificial intelligence, optimization algorithms, and expert systems in a probabilistic context in order to provide adaptive, constrained, and optimal decisions  \cite{borges2021strategic}. In real life, there are several works that use prescriptive analytics by integrating Machine Learning and Mathematical Optimization to build a unified framework on large scale data in different fields, such as marketing  budget allocation \cite{zhao2019unified}, online content recommendation \cite{agarwal2012personalized}, online advertisement \cite{xu2015smart}, email volume control \cite{zhao2018notification}, etc.

In the field of incentive allocation, the two-stage framework first predicts the customers' responses and then formulates the optimization problem based on business requirements, as proposed in some research works. Shen et al. \cite{shen2021framework} introduce a novel neural network to model personal promotion-response curves and optimize the ROI of various promotion campaigns. Li et al. \cite{li2020spending} focus on online coupon allocation with real-time intent detection. Albert et al. \cite{albert2021commerce} address cases with negative incremental net revenue. However, their prediction label and optimization objective are purchase propensity, instead of the purchase amount which better evaluates the campaign effect and is more challenging and complicated. Their research also does not mention the issue that the optimal solution based on Machine Learning results is being evaluated too optimistically \cite{ito2018unbiased}. We propose a novel offline evaluation method to overcome an intrinsic challenge in evaluating solutions.

\subsection{Customer Behavioral Characteristic Prediction}

To support the decision making, two types of modeling techniques for predicting customer behavioral characteristics are commonly applied: predictive modeling and uplift modeling. Predictive modeling predicts customer general future behaviors such as the number of transactions, churn propensity, and Lifetime Value (LTV); customers are then segmented based on the predictions for decision making. Uplift modeling predicts the customer response to treatments by establishing a causal link. To model and take advantage of the information brought by different treatments, uplift modeling introduces covariates transformation on treatments and response transformation \cite{gubela2017revenue}.

The first statistical model for predicting the LTV of customers is known as Buy 'Til You Die (BYTD). It models two processes of customer lifetime length and number of transactions by two separate parametric distributions. To tackle the issue of sparseness of data and incorporate the enormous customer data available in modern e-commerce scenarios, machine learning models are proposed. Malthouse et al. \cite{malthouse2005can} target LTV as a dependent variable in regression models directly. Because they use linear regression and LTV variance often increases with its mean, violating the assumption of homoscedasticity, they apply variance stabilizing transformations such as Box-Cox, square root and logarithm transformations. To tackle the LTV variance, some methods model the purchase propensity and purchase amount separately. Vanderveld et al. \cite{vanderveld2016engagement} and Chamberlain et al. \cite{chamberlain2017customer} use a two-stage random forest model framework. One model predicts whether or not the user will purchase in a given time window and the other predicts the dollar value for customers who are predicted to purchase.

DNN is also applied for behavioral characteristic prediction. Chamberlain et al. \cite{chamberlain2017customer} claim that DNN with enough number of neurons outperform Random Forest models. Wang el al. \cite{wang2019deep} integrate churn probability and LTV in one DNN structure by introducing zero-inflated lognormal (ZILN) loss. Shen et al. \cite{shen2021framework} use a DNN to learn user promotion response. They introduce a isotonic layer for a feature of promotion incentive level to satisfy monotonicity and smoothness.

\section{The Framework for Promotion Decision Making}

Our framework for promotion decision making consists of (1) customer response prediction including the direct purchase propensity and the enduring purchase amount, (2) optimization of incentive allocation, and (3) evaluation of allocation decision. Depending on the promotion design, we define the enduring effect as the impact in the period of four weeks, eight weeks or one quarter, during and after the promotion.  We do not consider a very long post-promotion period such as one year, because we believe a sole promotion can not have such a long-term effect and also apply subsequent promotions in a timely manner thus long-term effect is affected by multiple promotions.

This framework is applied to optimize the enduring effect on the purchase amount. The estimated cost per user is calculated as the product of the purchase propensity prediction during the campaign and the coupon value. Hence, the prediction targets of phase (1) consist of two parts: (a) predicting purchase propensity during the promotion to define the budget constraint and (b) predicting purchase amount during and post-promotion to formulate the optimization objective. Phase (2) is for the decision making, which optimizes marketing target (total purchase amount during and post-promotion) under the budget constraints by taking the prediction of phase (1) as the input parameters. After getting the result, the unbiased evaluation approach of the business metrics is applied in phase (3). The framework structure is shown in Figure \ref{fig:structure}. 

Section 3 explains the phase (2) and phase (3) and section 4 illustrates the architecture of the prediction model in phase (1). We summarize notations in Table \ref{table:notations}.

\begin{table}
\caption{Notations}
\label{table:notations}
\begin{center}
\begin{tabular}{ p{1cm} | p{6.5cm} } 
 \hline
 symbol & meaning \\
 \hline
$\boldsymbol{x}_{ij}$ & combined feature vector of the customer $i$ and incentive type $j$. \\ 
$s_{ij} $  & whether or not customer $i$ makes a purchase during the promotion when given incentive type $j$. \\ 
$y_{ij} $  & the enduring purchase amount of customer $i$ when given incentive type $j$. \\ 
$t_{ij}$ &  indicator variable representing whether or not incentive type $j$ is assigned to customer $i$ in RCT.  \\
$z_{ij}$ & decision variable representing whether or not to choose incentive type $j$ for customer $i$ in decision making optimization. \\
$f^\dagger(\boldsymbol{x}_{ij})$ & the predicted direct purchase propensity of customer $i$ when given incentive type $j$. \\
$f^{\S}(\boldsymbol{x}_{ij})$ & the predicted enduring purchase propensity of customer $i$ when given incentive type $j$. \\
$f^\ddagger(\boldsymbol{x}_{ij})$ & the predicted enduring purchase amount of customer $i$ when given incentive type $j$. \\
$c_{j} $   & the cost per acquisition of incentive type $j$. \\ 
$p_j$ & The proportion of customers in group $j$. \\
$M$ & the index set of incentives including zero-incentive. \\
$N$ & the index set of customers. \\
$B$ & the total campaign budget. \\
$Y$ & the total purchase amount. \\
 \hline
\end{tabular}
\end{center}
\end{table}
\subsection{Optimization Problem Formulation}
Suppose there are $|M|$ different types of incentives including zero-incentive and $|N|$ target customers in the promotion, we need to decide the type of incentive issued for each customer. Our objective is to maximize the total enduring purchase amount while keeping the cost under budget. The corresponding formulation is as follows:


\begin{equation}
\begin{aligned} 
  \textrm{maximize}   \quad &  \sum_{(i,j) \in N \times M} f^\ddagger(\boldsymbol{x}_{ij}) z_{ij} \\
  \textrm{subject to} \quad & \sum_{(i,j) \in N \times M}c_{j} f^\dagger(\boldsymbol{x}_{ij}) z_{ij} <= B, \\
  &\quad\:  \sum_{j\in M} \quad\:  z_{ij} = 1 \quad (i \in N ), \\
  &\quad\:   z_{ij} \in \{0,1\} \quad ((i,j) \in N \times M ).
\end{aligned} 
\end{equation}



\subsection{Evaluation}

In RCT, incentives are randomly allocated and are independent of customer behaviors, making it feasible to apply an unbiased estimation of the business impact under different incentive allocation strategies. In general, the value of the objective function for optimization is overestimated because of the variance of the model predictions \cite{ito2018unbiased}. To avoid the overestimation issue, we use the observed value of purchase amount to estimate decision making performance. Also, to assess the generalization capability of the solution, we conduct the evaluation with cross-validation. 

As we can not observe counterfactual customer response in RCT, we perform estimation by selecting customers whose assigned incentive type in our approach is the same as that in RCT (denoted as selected customers). We estimate $Y$ by the following three steps:

\begin{description}

\item[Step 1:] Calculate $\tilde{Y}_{j}$ which is the total purchase amount for each incentive type among selected customers:


\begin{equation}
\tilde{Y}_{j}=\sum_{i}y_{ij}z_{ij}t_{ij}\; \; (j \in M ).
\end{equation}

\item[Step 2:] As the customers are randomly assigned to different incentive groups in RCT, the unbiased estimation of $Y_{j}$ among customers assigned to incentive $j$ in the decision making strategy is calculated as: 

\begin{equation}
E(Y_{j})=\frac{\tilde{Y}_{j}\sum\limits_{i}{z_{ij}}}{\sum\limits_{i}{z_{ij}t_{ij}}}\; \; (j \in M ).
\end{equation}
\item[Step 3:]Based on the law of total expectation, the unbiased estimation of expectation of the total purchase amount $Y$ is calculated as: 

\begin{equation}
E(Y)=\sum_{j \in M}{p_{j}E(Y_{j})}.
\end{equation}
\end{description}



\begin{table*}
\caption{Evaluation results of the enduring purchase amount on COEFF, CORR, NRMSE, NMAE and the direct purchase propensity on AUC. For each metric, we highlight the best result in bold and underline the second best result.}
  \label{tab:metrics}
  \begin{tabular}{lcccccccccc}
    \toprule
    & \multicolumn{5}{c}{Spring Promotion} & \multicolumn{5}{c}{Winter Promotion} \\
    \cmidrule(lr){2-6}\cmidrule(lr){7-11}
    Method & COEFF & CORR & NRMSE & NMAE & AUC & COEFF & CORR & NRMSE & NMAE & AUC\\
    \midrule
    Two-Model approach \cite{lo2015predictive}  & 0.3277 & 0.2427 & 5.3117 & 1.4039 & 0.6498 & 0.3328 & 0.2330 & 3.9026 & 1.3090 & 0.6865 \\
    Two-Stage model \cite{chamberlain2017customer} & 0.4139 & 0.3179 & 5.1290 & 1.3229 & \textbf{0.6958} & 0.4379 & 0.3111 & 3.9152 & \underline{1.1966} & 0.7076 \\
    ZILN \cite{wang2019deep} & 0.4131 & \textbf{0.3288} & 10.7335 & 2.4891 & 0.6873 & 0.4646 & 0.3163 & 4.9672 & 1.2564 & 0.6839 \\
    \cmidrule(lr){1-11}
     CDEE (Ours) w/o CE Loss & \underline{0.4435} & 0.3036 & \underline{4.1070} & 1.3483 & 0.6832 & \underline{0.4778} & \underline{0.3335} & \textbf{3.4927} & 1.2259 & \underline{0.7112} \\
     CDEE (Ours) replace by L2 Loss & 0.3560 & 0.2239 & 4.1100  & \textbf{1.0161} & 0.6905 & 0.4177 & 0.3039 & 4.4867 & 1.2860 & 0.7025 \\
    \cmidrule(lr){1-11}
     CDEE (Ours) & \textbf{0.4822} & \underline{0.3230} & \textbf{3.9588} & \underline{1.2800} & \underline{0.6934} & \textbf{0.4924} & \textbf{0.3555} & \underline{3.5011} & \textbf{1.0040} & \textbf{0.7131} \\
    \bottomrule
  \end{tabular}
  
\end{table*}

\section{Customer Direct and Enduring Effect Model}

In this section, we elaborate on the customer direct and enduring effect (CDEE) model, which provides predictions of the direct purchase propensity and the enduring purchase amount of each customer.

We apply a deep neural network (DNN) for prediction due to its performance and flexibility. DNNs have achieved great success in computer vision, natural language processing, video/speech recognition, etc. They also show competitive performance on tabular data \cite{gorishniy2021revisiting}. Moreover, we can design customized hybrid loss functions in DNNs, which allows us to build a loss function to estimate the direct and enduring response for multiple treatments together in one unified model. 

In general, purchase amount regression is more difficult than purchase propensity prediction due to the distribution of the target variable. The distribution of purchase amounts is multimodal and highly skewed. There is a large proportion of customers who will not purchase, resulting in a zero-inflated distribution. Also, among the customers who do purchase, the purchase amount starts from some certain value, leaving a gap between the rest of the distribution and zero. Furthermore, the purchase amounts of different customers differ by several orders of magnitude. L1 loss and L2 loss, the typical regression losses, over-penalize high-value predictions and are sensitive to outliers. 

The compound Poisson–Gamma distribution, a subclass of the Tweedie distribution \cite{tweedie1984index}, models a positive mass at zero and a continuous response, which makes it suitable for fitting the purchase amount distribution. Given by $\rho \in (1,2)$ and $\mu$ > 0, the probability density function (pdf) of the distribution on variable $y$ is:

\begin{equation}
  f_{\rm{Tweedie}}(y|\mu,\phi,\rho)=\exp\Big\{{\frac{1}{\phi}\big(y\frac{\mu^{1-\rho}}{1-\rho}-\frac{\mu^{2-\rho}}{2-\rho}\big)}\Big\}a(y,\phi,\rho),
\end{equation}
where $\mu$, $\phi$ and $\rho$ are the mean, the dispersion parameter and the index parameter of the distribution respectively.  $a(y,\phi,\rho)$ is a normalizing constant \cite{yang2018insurance}. We omit subscripts in this section for simplicity. We take $\mu$ as the estimator of the purchase amount (denoted as $\hat{y}$) and derive the loss function through the negative log-likelihood as:

\begin{equation}
  L_{\rm{Tweedie}}(y, \hat{y}, \rho)=-y\frac{\hat{y}^{1-\rho}}{1-\rho}+\frac{\hat{y}^{2-\rho}}{2-\rho}.
\end{equation}

We propose a layered hybrid loss in the DNN model to jointly learn both direct purchase propensity and enduring purchase amount predictions. We construct a multi-task learning structure to exploit commonalities across tasks of direct and enduring response learning, and purchase propensity and amount learning. As shown in Figure \ref{fig:structure} (b), in the lower embedding layer, we employ a cross-entropy loss on the direct purchase propensity. In the higher embedding layer, we employ a Tweedie loss on the enduring purchase amount. We notice that by using the Tweedie loss on its own, the predictions form a unimodal distribution by predicting all the customers to have a non-zero purchase amount. To better identify non-purchase customers, we add a cross-entropy loss for the enduring purchase propensity in the middle layer. We combine the loss functions as:

\begin{equation}
  \label{eq:loss}
  \begin{aligned}
  L =& w_1 L_{\rm{Tweedie}}\big(y;f^\ddagger(\boldsymbol{x}),\rho\big) \\ 
  &+w_2 L_{\rm{CrossEntropy}}\big(\mathbb{1}_{\{y>0\}};f^{\S}(\boldsymbol{x})\big)\\ 
  &+w_3 L_{\rm{CrossEntropy}}\big(s;f^\dagger(\boldsymbol{x})\big),
  \end{aligned}
\end{equation}
where $w_1$, $w_2$, and $w_3$ are the weights for each loss function respectively.  In contrast to the typical solution of training two models for purchase propensity and purchase amount separately, the CDEE model integrates both of them into a unified model by applying Equation (\ref{eq:loss}). We enhance both direct and enduring response training by applying multi-task learning. On top of that, to improve the regression performance for the enduring purchase amount, we add a cross-entropy loss on the enduring purchase propensity.


\section{Experiment}



\subsection{Randomized Control Trial}

We conduct evaluation on the RCT data of two promotions in Mercari. To eliminate seasonal effects on the evaluation, we choose two promotions running in different time spans, one in spring (denoted as spring promotion) and the other in winter (denoted as winter promotion).

The goal of the two promotions is to improve customer retention by distributing coupons. We target customers who have not made a purchase for a certain period and deem them as potential churn customers. Coupons have different types including discounts and rebates, and different values. There are six types of coupons in the spring promotion and five types of coupons in the winter promotion. A no-coupon (control) group is also included in both promotions. The coupon validity period is five days. To evaluate the promotion performance considering a post-promotion dip in purchase, we evaluate the business impact on the purchase amount in four weeks after issuing coupons. The number of target customers in each promotion is approximately 650,000 and 600,000, respectively.

\subsection{Evaluation of Direct and Enduring Response Predictions}

For the model implementation, we set the embedding layer units of the incentive types as $\lfloor |M|^{0.25} \rfloor +1$ and hidden layers with 1024, 1024, 512, and 16 units. We use Adam optimizer \cite{kingma2014adam}, set the learning rate as 0.0002 and reduce the learning rate by 0.1 once learning stagnates. We set the batch size as 1024, and use an early stop with a patience of 10 epochs. We also apply dropout with a rate of 0.2. We set $w_1$, $w_2$ and $w_3$ as 10, 1, and 2 respectively. Finally, we set $\rho$ as $1.5$.

Features are extracted from a variety of resources. We create recency and frequency features out of transactions and interactions such as views, comments, likes, etc. And we additionally use monetary value features extracted from transactions. For each feature of frequency and monetary value, we aggregate them into long-term and short-term. We additionally take the incentive types as categorical features and encode them as embeddings.

We evaluate the model performance of the prediction of direct purchase propensity and enduring purchase amount respectively. We apply Area Under Curve (AUC) to evaluate the direct purchase propensity prediction. For the evaluation of enduring purchase amount prediction, we evaluate on both error metrics and discrimination metrics. Following \cite{xing2021learning}, we apply Normalized Root Mean Square Error (NRMSE) and Normalized Mean Average Error (NMAP) as error metrics. Considering that the normalized error metrics amplify large prediction errors and over-emphasize high-value purchase amounts, to make evaluation robust to those data points, we evaluate model discrimination performance by evaluating the ranking of the predictions. We apply Spearman's correlation (CORR) because a normal distribution is not assumed in our data. Since the Gini coefficient is frequently used in economics, we also use a normalized Gini coefficient (COEFF) \cite{wang2019deep} to quantify model discrimination. The higher the COEFF, CORR, and AUC scores, and the lower the NRMSE and NMAE, the better the predictions of the model.

We compare the CDEE model with two customer response methods: the two-model approach \cite{lo2015predictive} and the two-stage model \cite{chamberlain2017customer}. We also compare the CDEE model with the zero-inflated lognormal (ZILN) model \cite{wang2019deep} which focuses on fitting the highly skewed distribution of the purchase amount. We evaluate all the methods with five-fold cross-validation on the two promotion datasets.

As shown in Table \ref{tab:metrics}, with respect to the discrimination metrics and error metrics, the CDEE model performs well on both prediction tasks. The ZILN model performs well on discrimination metrics but performs weakly on error metrics, which might be caused by under-fitting the low-value data when trying to fit the log-normal distribution. Judging by the AUC for the direct purchase propensity, compared with the other DNN model ZILN, the CDEE model gets higher scores, showing that multi-tasking learning also benefits propensity prediction. The two-stage model is a tree-boosting-based model. While in general tree boosting algorithms tend to perform the best when it comes to propensity prediction on tabular data, We find the CDEE model has a comparable performance with it.

To check the effectiveness of each component in the CDEE model, we conduct an ablation study by removing enduring CE loss and replacing Tweedie loss with L2 loss. The results show that both adding CE loss and applying Tweedie loss improve the model performance on the two datasets.

\subsection{Business Performance Evaluation}

As per business requirements,  the goal of the promotions is to maximize the enduring lift ROI while keeping the cost under the budget. Thus, lift purchase amount (LPA) during and post promotion period, and the incentive cost are key factors. LPA is the difference between the purchase amount of the treatment group and the control group, and is used to measure the four-week enduring effect. The value of each type of coupon is constant. Since LPA is affected by both the incentive allocation method and the incentive cost, we compare LPA of different methods by fixing the cost and compare LPA of each method at different costs. 
 
\begin{figure}
  \includegraphics[width=0.45\textwidth]{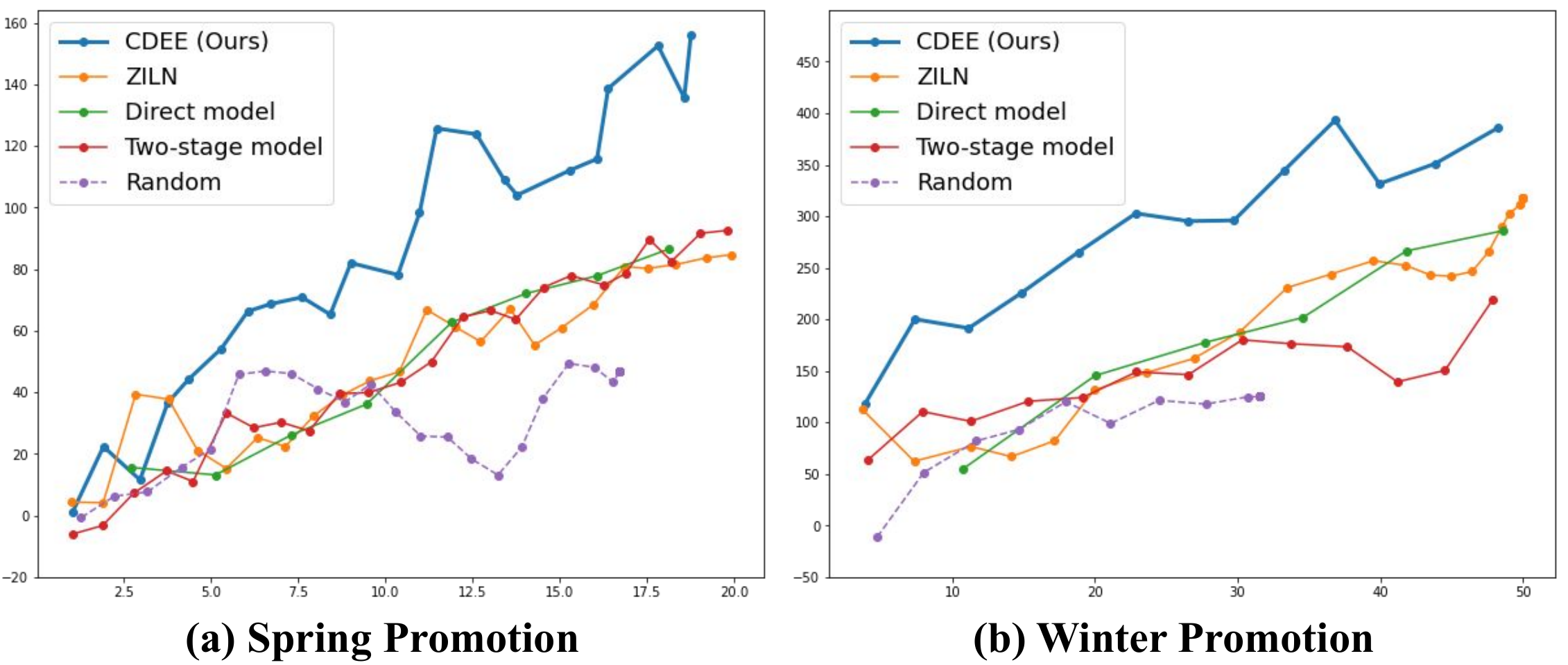}
  \caption{Evaluation results on comparing the CDEE model with other methods under different costs on the two promotion datasets, where x-axis represents the incentive cost, and y-axis represents the LPA.}
  \Description{}
  \label{fig:experiments}
\end{figure}

We evaluate the business performance of four methods. To compare the predictive model performance, we gather predictions from the CDEE model, the second-best models in Section 5.2: ZILN and Two-Stage model, and a random prediction. We apply the same optimization process to these predictions. To demonstrate that the performance of the enduring effect can not be guaranteed when only modeling the direct response, we train a CDEE model by changing all the labels to direct response (denoted as Direct model). Figure \ref{fig:experiments} shows the LPA given by the different incentive costs. We observe that the CDEE model outperforms the other predictive models in the business metrics. Direct model has significantly lower LPA compared to the CDEE model. It indicates that the direct effect is not proportional to the enduring effect. Modeling the enduring effect is necessary when we aim to improve the customer retention and loyalty. 

We also observe that as the budget amount increases, the expected LPA increases for all the models. This is in line with the intuition that under the same conditions, when the budget increases, customers receive larger value of coupons and are hence willing to place more or higher value orders. However, this intuition is not applicable to the random prediction method. The random prediction method exhibits fluctuation in the LPA, especially in the spring promotion dataset, since optimizers act as “error maximizers” \cite{ito2018unbiased} when the prediction contains large errors. As the prediction value is randomly generated, we can not issue the optimal coupon for each customer even if we have enough budget. This causes the upper bound of the incentive cost and LPA in the random prediction method are much lower than the other methods.  

\section{Conclusion}

To create promotions that aim to build long-term customer engagement and boost customer loyalty, we target both during and post-promotion period effect. To make a better decision for multiple treatment promotions considering the contribution of the customer purchase amounts and the promotion enduring effect, we propose a framework by modeling customer direct and enduring response. Our framework consists of direct and enduring response predictions, incentive allocation decision making and business impact evaluation. To target the highly skewed distribution of purchase amount, we propose the CDEE model. We compare our method with benchmarks using two promotions in Mercari and achieve significantly better results on both model evaluation and business metrics.

\printbibliography
\appendix
\end{document}